# An Introduction to Communication Efficient Edge Machine Learning

*Qiao Lan, Zezhong Zhang, Yuqing Du, Zhenyi Lin, and Kaibin Huang*[1]

## Abstract

In the near future, Internet-of-Things (IoT) is expected to connect billions of devices (e.g., smartphones and sensors), which generate massive real-time data at the network edge. Intelligence can be distilled from the data to support next-generation AI-powered applications, which is called *edge machine learning*. One challenge faced by edge learning is the communication bottleneck, which is caused by the transmission of high-dimensional data from many edge devices to edge servers for learning. Traditional wireless techniques focusing only on efficient radio access are ineffective in tackling the challenge. Solutions should be based on a new approach that seamlessly integrates communication and computation. This has led to the emergence of a new cross-disciplinary paradigm called communication efficient edge learning. The main theme in the area is to design new communication techniques and protocols for efficient implementation of different distributed learning frameworks (i.e., federated learning) in wireless networks. This article provides an overview of the emerging area by introducing new design principles, discussing promising research opportunities, and providing design examples based on recent work.

## 1. Introduction

While oil was the world's most valuable resource a century ago, the role has been assumed by data in the digital era where Internet companies (e.g., Google and Amazon) dominate the economy. The extreme popularity of smartphones and Internet is driving the continuous generation of abundant and ubiquitous data. Practically any activity of a mobile user leaves a digital trace in the form of raw data. For instance, a future self-driving car is estimated to generate 100 gigabytes per second. Leveraging recent breakthroughs in *artificial-intelligence* (AI) techniques, the massive available data can be distilled into intelligence for supporting new applications. Specifically, intelligence algorithms can predict the preferences of a customer, the needed servicing of an engine, and the health risk of a person. The possibility of obtaining such intelligence from abundant distributed data inspires people to envision ubiquitous computing and ambient intelligence in the near future.

Next-generation wireless networks (e.g., Internet-of-Things) are expected to connect tens of billions of edge devices (e.g., smartphones and sensors). Massive real-time data will be continuously generated at the network edge. This makes the network a gigantic data source for intelligence distillation. However, the aggregation of distributed data for training AI models in the cloud faces two main challenges among others. First, typical sensing data (from e.g., camera and LIDAR) are high-dimensional and the transportation of a large volume of data across a network will cause traffic congestion. Second, the long round-trip latency between the cloud deep in the network and user devices at the edge makes it difficult for a machine to learn from or react to real-time events in mission critical applications requiring tactile-speed responses. The attempts to tackle the challenges have motivated both the industry and academia to develop technologies for deploying AI algorithms at the network edge, called *edge AI* [1, 2]. Edge learning refers to the first phase of edge AI, training of an AI model, while the other phase is the application of the trained model to inference. The pushing of learning towards the edge will allow rapid access to the enormous real-time mobile data and reduce traffic congestion in backhaul networks.

[1] Q. Lan, Z. Zhang, Y. Du, Z. Lin and K. Huang are with the Department of Electrical and Electronic Engineering, The University of Hong Kong, Hong Kong. Corresponding author: K. Huang (email: huangkb@eee.hku.hk).

5G networks are required to support numerous services including mobile broadband, machine-type communication, and ultra-reliable low-latency communication. The multiplicity of services causes a massive number of devices accessing a network and results in the average speed (50-100 Mb/s) far below the peak rate (several Gb/s) due to the scarcity of radio resources. Further loading the networks with the edge learning service will inevitably exacerbate the situation, creating a communication bottleneck. Overcoming the bottleneck by communication efficient designs is a main theme of edge learning research. The state-of-the-art wireless technologies are insufficient. Their designs attempt to achieve the goals of communication reliability and data-rate maximization, and adopt the approach of "communication-computing separation". These goals are not directly those of edge learning which aims at fast and accurate intelligence acquisition. Attempts to achieve the latter have led to the emergence of a new area, called communication efficient edge learning. In this area, researchers seek to seamlessly integrate communication and learning for unleashing the full potential of edge learning. This article provides an introduction to the area by describing the background, discussing the design principles, identifying promising research opportunities as well as presenting some design examples.

The remainder of the article is organized as follows. The area of edge machine learning is overviewed in Section 2. Three directions for communication efficient edge learning, namely multiple access, data preprocessing and radio resource management, are introduced in Sections 3 to 5, respectively. Concluding remarks are provided in Section 6.

## 2. Overview of Edge Machine Learning

As AI is expected to enable many next-generation mobile applications, machine learning will be common operations in 5G networks. As illustrated in Fig. 1, the computing resources in the networks that can be leveraged for learning can be grouped into three layers: *central cloud*, *edge cloud*, and *on-device* resources, which are located at data centers, network edge (e.g., base stations, edge servers, and micro-grids), and mobile devices, respectively. As also shown in Fig. 1, the three layers are differentiated by their computing capacities, namely computing speeds from high to low and storage from large to small, and their communication capacities, namely end-to-end latency from long to short and bandwidth from small to large. Learning tasks with different requirements (work load, latency, and bandwidth) will be distributed over different computing layers of the network. In particular, tasks requiring the training of a huge AI model using a massive dataset should be executed in the central cloud; those with the training of a low-complexity model but a stringent latency requirement can be performed on device. Edge cloud strikes a balance between central cloud and on-device computing in forms of computation and communication capacities. Edge learning, or learning using the edge cloud, has numerous advantages described as follows.

- **Low latency**: The learning latency comprises communication (propagation and multiple access) and computation latency. Edge servers are typically located in proximity (e.g., tens of meters) to edge devices, the data sources. In contrast, the distances between a central cloud and devices can range from tens of kilometers to those crossing continents. Moreover, the multi-access channel and the computation resources at an edge server are shared by a much smaller number of users compared with those at a cloud server. For this reason, the latency of edge learning is much shorter than that of central-cloud learning. This makes it possible for the former to support high-mobility applications such as UAV control or tactile-speed applications such as Internet-of-Skills.

- **Context and location awareness**: Compared with central-cloud learning, the proximity of edge learning to users allows a machine to learn the users' behaviors, locations, and environments. This endows on the trained AI models context and location awareness. Consequently, they can provide intelligent context-aware services to end users. For instance, based on their real-time behaviours and locations, machines can order and download contents onto users' devices in advance or dynamically manage their daily schedules.

- **Data safety**: Central clouds are hosted in remote public data centers, such as the Amazon EC2 and Microsoft Azure, which are susceptible to attacks for their being concentrated sources of private information. Edge learning consumes users' data at the network edge without their uploading to data centers. This reduces their risk of leakage. Some edge learning frameworks, such as federated learning discussed in the sequel, preserve user privacy by keeping data on devices and requiring them to only upload model updates to edge servers that do not reveal the content of data. Furthermore, many edge clouds are privately owned by enterprises where strong security measures are implemented against illegitimate external access.

- **Reduced network congestion**: A typical dataset for training an AI model is massive e.g., comprising millions of photos. The transportation of many such datasets across a backhaul network can severely increase the current traffic congestion. The issue can be addressed by edge learning where data is generated and consumed only at the network edge without affecting the backhaul network.

- **Leverage of distributed data and computing resources**: Edge learning typically involves an edge server distributing the computation load over many devices or aggregating distributed data from them. Compared with on-device learning, edge learning is capable of leveraging distributed computing resources and a much larger and diversified dataset for training a more complex AI model with a higher accuracy.

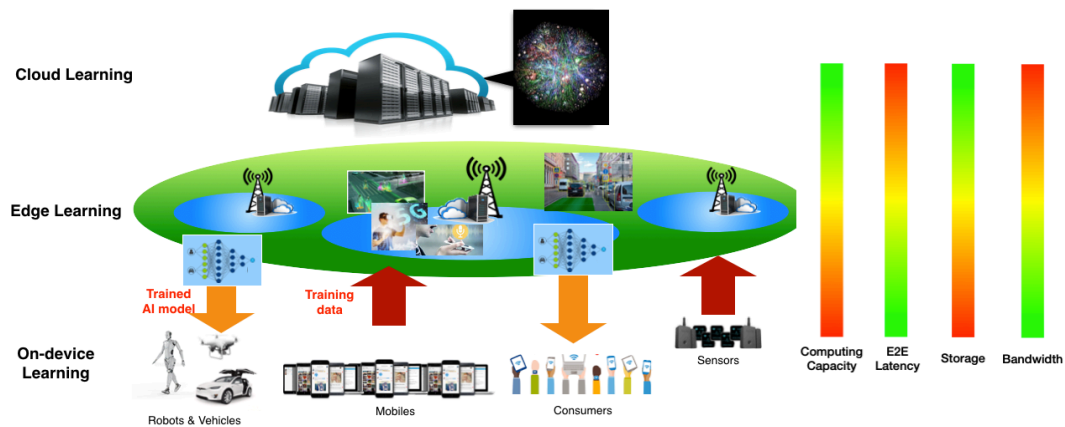

Fig. 1. Layered machine learning in a 5G network with different capacities.

A large collection of generic open datasets are available in the cloud ranging from conversations to images of annotated objects (e.g., cars, humans, and animals). The datasets allow the training of generic AI models (e.g., speech and image classification) that provide a basis for edge learning. The typical procedure for edge learning is as illustrated in Fig. 2 and described as follows. The learning at an edge server starts with downloading an initial model from the generic AI-model library in the central cloud. Then the model is improved at the server for a specific application (e.g., autonomous driving) using domain, context or location specific data acquired from devices. The generic model can be modified using one of many available machine learning techniques including transferred learning, reinforcement learning, or active learning. Upon the completion of learning, the trained model customized for a specific application can be then downloaded onto a device for inference, interpretation, vision, sensing, and control (see Fig. 2).

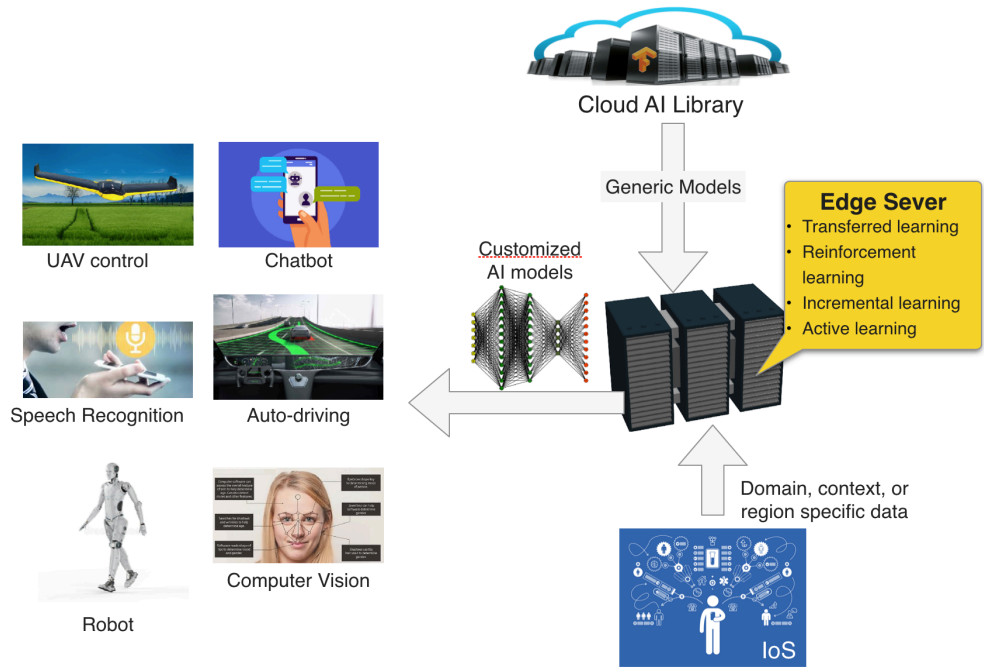

*Fig. 2. Typical procedure of edge machine learning.*

## 2.1 Communication-Efficient Edge Learning

Two key missions that have been driving the advancements in 5G technologies are gigabit access and tactile response time. The first mission of gigabit access aims at achieving transmission rates of several gigabit-per-second. The high rates make it possible to implement data-hungry new applications such as *ultra high definition* (UHD) video streaming and *virtual reality* (VR). On the other hand, the other mission of tactile response time aims at achieving fast network response time that can approach the human reaction speed (within several milliseconds). Without such speeds, safety and accuracy cannot be guaranteed for latency sensitive applications such as autonomous driving, cloud controlled robotics, and industrial automation. Though 5G networks are being deployed, the state-of-the-art technologies have not yet achieved the two missions. Measurements have shown that the average access speed is in the range of 50-100 Mb/s that is an order-of-magnitude slower than the targeted rate. The main obstacle is that the radio access network is required to support the multiple access by a massive number of users and IoT devices. On the other hand, the present achievable communication latency ranges from tens to hundreds of milliseconds. Tactile-speed response is difficult to achieve as computing (at both devices and base stations), protocols (e.g., admission and routing), and round-trip propagation all incur delay and suppressing the total latency below ten milliseconds is challenging. As they are missions not yet accomplished, there is no doubt that gigabit access and tactile response will continue to be the goals of 6G development. As mentioned, motivated by the availability of massive data at the edge and the power of AI, a new and key 6G mission will be to realize the vision of ubiquitous computing and intelligence, providing a platform for supporting next-generation intelligent applications. The mobile edge computing architecture introduced in the 5G standardization has already laid a foundation for materializing this 6G mission. In this architecture featuring network virtualization, the virtualization layer aggregates computation resources that are geographically distributed such that they form a virtual cloud on which different cloud applications can be deployed. Realizing the 6G mission has motivated both the industry and academia to conduct extensive research and development on edge computing and learning. However, the full potential of these technologies cannot be unleashed without achieving gigabit access and tactile response. In particular, they are crucial for enabling edge learning to efficiently exploit massive distributed real-time data and quickly respond to "black-swan" type events. As computing speeds are increasing rapidly, the scarcity of radio resources for data intensive communication between edge servers and devices creates a communication bottleneck for edge learning.

Overcoming the communication bottleneck calls for a fundamental change in design principles and approaches. Originating from Claude Shannon's pioneering work, the traditional approaches decouple communication and computing (or learning in the current context) and focus on rate maximization or communication reliability. As the result, wireless technologies developed based on this approach are inefficient in tackling the edge-learning challenges. They overlook the fact that for edge machine learning, we are interested in computing some overall function of the distributed data rather than obtaining the data from individual user devices. For example, federated learning discussed in the sequel is based on *stochastic gradient descend* (SGD) and requires the server to acquire only the average of transmitted data (gradients) instead of the full knowledge of all gradients. By overlooking this fact, the conventional multiple-access schemes based on orthogonalization treats interference as being harmful. As a result, the required radio resources linearly increase with the number of accessing devices. In the context of edge learning, it is possible to harness interference for fast multiple access (see Section 3). As another example, based on the rate-maximization approach, different data are treated as having equal importance but from the active learning perspective, some are more important than others. Without considering this fact, the air interface may be congested by the transmission of data that are not useful for edge learning.The drawbacks of the traditional approaches call for the development of the new design principles and techniques, called communication efficient edge learning, based on the following new principle.

---

**Principle of Communication-Efficient Edge Learning**

To jointly design distributed machine learning and wireless communication techniques for the objective of fast and accurate intelligence distillation at edge servers from distributed data at edge devices.

---

In Sections 3-5, we shall discuss specific research directions in communication efficient edge learning and provide concrete examples to illustrate this paradigm shift, which cover key communication aspects including multiple access, data compression and resource allocation.

## 2.2 Two Paradigms of Edge Learning

There exist two main paradigms of edge learning: *centralized edge learning* and *federated edge learning*, introduced as follows.

**Federated Edge Learning**: The paradigm targets open networks (e.g., WiFi and cellular networks) where user privacy is a critical concern. The paradigm preserves user privacy by avoiding data uploading and distributing learning at both servers and edge devices, which is coordinated using wireless links [3-5]. There exist two schemes for federated learning featuring different types of updates by edge devices. For one scheme, the edge devices transmit to the server their local models, whose average replaces the global model. The other scheme is based on the method of SGD, where devices transmit local gradient vectors and their average is applied to update the global model using SGD. As illustrated in Fig. 3(a), for either scheme, the learning procedure iterates between 1) the broadcast by the server the global model for updating local models and 2) transmission by edge devices for updating the global model. The iteration continues until the global model converges. Each iteration is called a *communication round*. For federated learning, the concern of communication overhead has also motivated researchers to develop algorithms for selecting only a subset of devices with important updates for transmission. The theme is called active (model) update acquisition. There exist two importance metrics corresponding to the two mentioned schemes. One is *model variance* which indicates the divergence of a particular local model to the average of all local models. The other is *gradient divergence* that reflects the level of changes on the current gradient update w.r.t. the previous one. Using these metrics to schedule updating devices, a number of so called "lazily updating" algorithms have been designed for communication-efficient federated learning [6, 7]. Another approach for reducing communication overhead is to compress gradient vectors by exploiting their sparsity in significant elements [8, 9].

**Centralized Edge Learning**: The paradigm targets closed sensor networks (in e.g., enterprises, home or factories) where privacy is not an issue. For this paradigm as illustrated in Fig. 3(b), learning is performed only at the server and training data is directly acquired from edge devices by wireless transmission [10]. Consider distributed learning by direct data acquisition. To improve communication efficiency, it is natural to acquire only data samples that are "important" for learning. Two data-importance metrics from the area of active learning are *data uncertainty* and *data diversity* [11]. Data classification is an important topic in machine learning that has a wide-range of applications ranging from computer vision to information retrieval. Targeting classification, active learning largely concerns the selection of data samples form a large unlabelled dataset for manual labelling. The labelled data can be then used to train a classifier model by supervised learning. Considering a single data sample, its informativeness for learning can be quantified by data (classification) uncertainty, referring to how confident it can be correctly classified using the current model [12]. On the other hand, considering a batch of samples, the total informativeness depends not only on uncertainty of individual samples but also on their *diversity*, referring to non-overlapping information [13].

The differences between the two paradigms are summarized in Table 1. Both paradigms are considered in designing techniques for communication efficient edge learning in the following sections.

Table 1. Comparison between Federated Learning and Centralized Learning

|  | **Federated Learning** | **Centralized Learning** |
| --- | --- | --- |
| **Application Scenario** | Open networks | Closed sensor networks (enterprises, homes, factories) |
| **Communication Situation** | Multiple rounds of model uploading and downloading | One shot data uploading |
| **Data Privacy** | Data kept local and privacy preserved | Data uploaded |

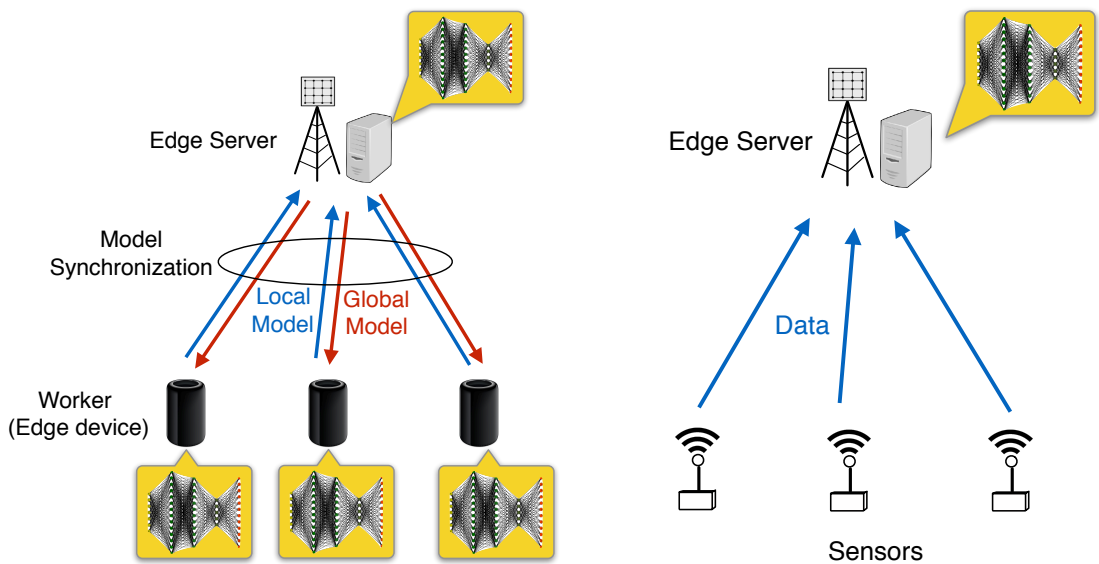

*Fig. 3. Two key paradigms of edge learning: (a) (left) federated edge learning and (b) (right) centralized edge learning.*

## 3. Multiple Access for Edge Learning

### 3.1 Recent Trends and Opportunities

Consider the paradigm of federated edge learning introduced in Section 2 and illustrated in Fig. 3(a). The learning task is to train an AI model typically comprising millions to billions of parameters. The iterative learning process usually lasts hundreds of communication rounds. In each round, multiple edge devices upload locally updated models (or stochastic gradients of the same size) to the edge server over a multiple access channel. The long training duration, the multi-access by many devices and the large volume of data uploaded by each device can cause severe traffic congestion in the air interface. The difficulty cannot be overcome by the conventional orthogonal-access schemes including *orthogonal frequency-division multiple access* (OFDMA), *time time-division multiple access* (TDMA) and *code division multiple access* (CDMA). These rate-maximization schemes are inefficient for supporting local-model aggregation in federated learning. They required the radio resources to linearly increase with the number of devices and the volume of data.

The shift of the objective from the conventional sum rate maximization to functional computation, namely model/gradient aggregation, calls for a change in the design principle for multi-access schemes. We propose that the **new principle** should be:

> *A multi-access scheme for federated edge learning should harness interference to reduce the required radio resources (time, spectrum, antennas, and power) consumed in the learning process.*

Based on the principle, a new class of multiple-access techniques called *Over-the-Air Computation* (AirComp) have been recently developed to support efficient aggregation for federated edge learning [14-18]. Instead of treating interference as being harmful, AirComp harnesses interference to realize functional computation over a multi-access channel. To be specific, AirComp exploits the waveform superposition property of the multi-access channel. As a result, the simultaneous transmission of analog modulated models/gradients by devices allows the AP (or server) to directly receive the desired model/gradient average. The deployment of AirComp for aggregation in a federated edge learning system is illustrated in Fig. 4. AirComp is a simultaneous access scheme and thus the required radio resources is in theory independent of the number of devices.

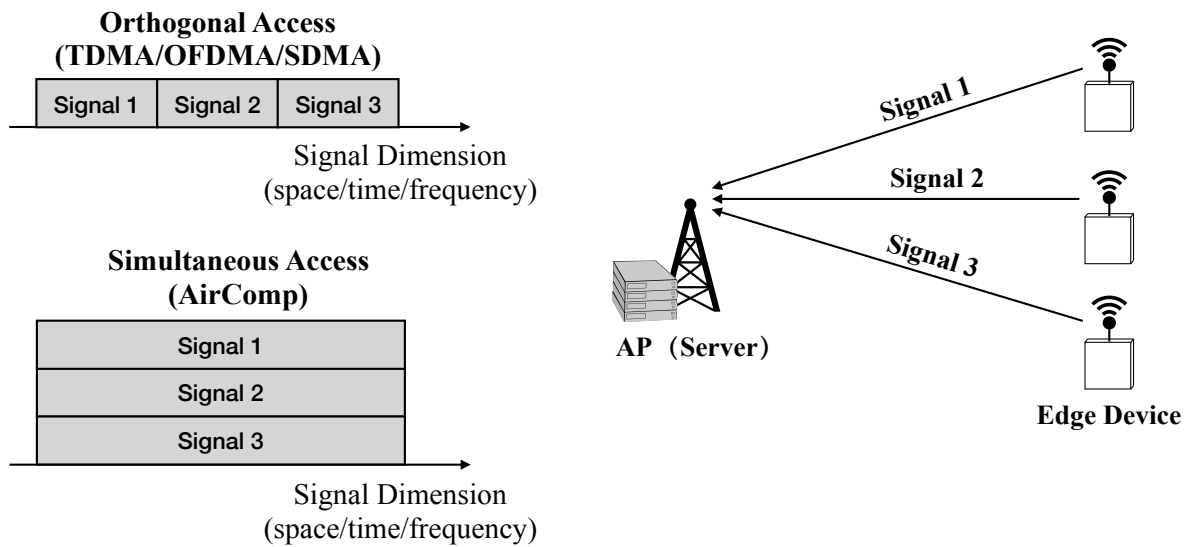

*Fig. 4. Comparison between the conventional "transmit-then-compute" with interference management approach and AirComp (simultaneous transmission) approach.*

Despite some initial progresses, AirComp for federated edge learning is in a nascent stage where there lie many opportunities. Some of them are described as follows.

1) **Multi-cell AirComp**: One topic that has not been explored is AirComp for a multi-cell system. New techniques for inter-cell interference management need be designed to maximize the number of aggregated signal streams in the system. In particular, the classic interference-alignment theory can be redeveloped to account for AirComp operations such as *signal alignment in magnitude* and the learning performance metric including convergence rate and accuracy.

2) **AirComp with digital modulation and coding**: The basic AirComp design assumes linear analog modulation while digital modulation and coding are widely used in practical wireless systems such as 3GPP. One important direction is to design AirComp with digital modulation and coding for implementing federated edge learning in wireless networks. This is possible as the popular modulation scheme, QAM, is a type of amplitude modulation (albeit with discretisation errors) like linear analog modulation. Moreover, the integration of AirComp with specific coding schemes such as lattice coding has been explored in prior work [19, 20].

3) **Channel estimation and feedback for imperfect AirComp**: Imperfect *channel state information* (CSI) at the server or devices results in AirComp errors that affect learning accuracy. It calls for the design of efficient schemes for channel estimation and feedback exploiting the AirComp protocol and architecture to reduce the overhead for pilot transmission and CSI feedback. Some initial results are reported in [14, 15]. Furthermore, it is important to quantify the effects of AirComp errors on the performance of edge learning. This facilitate the resource allocation for channel training and the design of new techniques for improving the learning robustness against CSI errors.

4) **AirComp for highly heterogeneous devices**: Current AirComp schemes require strict synchronization among all the participating edge devices. The synchronization requirement may increase the multi-access latency in a system with high heterogeneity in devices' computing speeds and propagation latency, thereby compromise the key advantage of AirComp. For such scenarios, it is necessary to develop AirComp technologies for coping with the high heterogeneity. User selection can be adopted to exclude extremely slow devices and prevent them from becoming a bottleneck for AirComp. On the other hand, AirComp can be assisted by radio resource management that to some extent can equalize the heterogeneity.

5) **MIMO AirComp**: The deployment of antenna arrays support spatially multiplexing of AirComp streams to reduce the duration of communication round in federated edge learning. One issue is the joint design of receive aggregation beamformer at the server and transmit beamformers at devices under the criterion of minimum AirComp error. The issue has been addressed in some initial work [15]. Many other interesting issues warrant further investigation. Based on the concept of *age-of-information* (AoI), AoI-aware aggregation beamforming can be designed for asynchronous federated learning where uploaded local models are weighted according to their "ages" (since last uploading) to balance the tradeoff between exploitation of distributed data diversity and the performance degradation of outdated updates. On the other hand, similar to the known result in information theory, there exists a fundamental tradeoff between spatial multiplexing and diversity from the new perspective of learning performance. Such a tradeoff has not been studied.

## 3.2 Design Example

We discuss one design example of AirComp, namely MIMO AirComp, for federated edge learning. The example is based on the application of the generic MIMO AirComp design in [15] to federated edge learning. Consider the federated edge learning system in Fig. 3(a) where antenna arrays are deployed at both the AP and devices. The arrays create MIMO channels for supporting spatial multiplexing of AirComp streams. A high-dimensional local model (or gradient) to be transmitted by a device is divided into blocks. Each block is transmitted as a vector symbol that is modulated using linear analog modulation to facilitate AirComp. Precoding and receive beamforming are applied at devices and AP, respectively. They are jointly designed to minimize the errors in the received AirComp data (averaged model/gradient). The block diagram illustrating the AirComp operations is shown in Fig. 5.

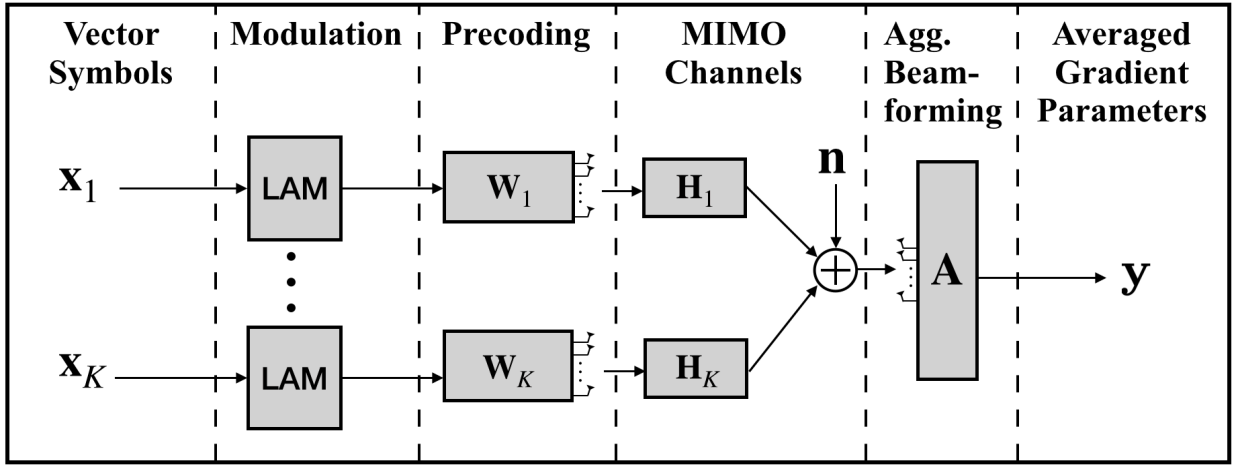

*Fig. 5. Federated learning system with MIMO AirComp.*

Let $M_r$ and $M_t$ denote the sizes of the receive and transmit arrays, respectively, and $N$ the number of spatially multiplexed AirComp streams. The received vector symbol at the AP can be written as

$$\mathbf{y} = \mathbf{A}^H \sum_{k=1}^{K} \mathbf{H}_k \mathbf{W}_k \mathbf{x}_k + \mathbf{A}^H \mathbf{n}$$

where $\mathbf{A}$ is a $M_r \times N$ aggregation beamformer, $\mathbf{H}_k$ is the $M_r \times M_t$ MIMO channel of the $k$-th device, $\mathbf{W}_k$ is the $M_t \times N$ precoder, $\mathbf{x}_k$ is the $N \times 1$ vector symbol (block of gradient parameters), and $\mathbf{n}$ is the white Gaussian noise vector. Zero-forcing precoding is applied to inverse MIMO channels so as to align multiuser signals at the AP to facilitate AirComp. To this end, $\mathbf{H}_k$ can be decomposed as SVD as

$$\mathbf{H}_k = \mathbf{U}_k \begin{bmatrix} \lambda_k^{(1)} & 0\cdots & 0 \\ 0 & \lambda_k^{(2)}\cdots & 0 \\ \vdots & \vdots \ddots & \vdots \\ 0 & 0\cdots & \lambda_k^{(M_t)} \\ \vdots & \vdots \ddots & \vdots \\ 0 & 0\cdots & 0 \end{bmatrix} \mathbf{V}_k^H.$$

Assuming sufficient transmission power, the zero-forcing precoder is given as

$$\mathbf{W}_k = \mathbf{V}_k \begin{bmatrix} 1/\lambda_k^{(1)} & 0\cdots & 0 \\ 0 & 1/\lambda_k^{(2)}\cdots & 0 \\ \vdots & \vdots \ddots & \vdots \\ 0 & 0\cdots & 1/\lambda_k^{(M_t)} \end{bmatrix}.$$

As a result, the received vector symbol reduces to

$$\mathbf{y} = \sum_{k=1}^{K} \mathbf{A}^H \mathbf{U}_k \mathbf{x}_k + \mathbf{A}^H \mathbf{n}.$$

The above expression implies that to minimize the AirComp errors (or equivalently maximize the receive SNRs), an optimal aggregation beamformer $\mathbf{A}$ needs to balance its subspace distances to the left eigen-spaces of all individual MIMO channels. The error minimization problem can be approximately formulated as the following optimization problem [15].

(P1) $$\min_{\mathbf{A}} \sum_{k=1}^{K} d_{\text{P2}}^2 \left( \mathbf{U}_k, \mathbf{A} \right)$$
$$\text{s.t.} \quad \mathbf{A}^H \mathbf{A} = \mathbf{I}$$

In Problem P1, the constraint indicates that $\mathbf{A}$ is a tall unitary matrix typical for receive beamforming and $d_{\text{P2}}\left(\mathbf{U}_k, \mathbf{A}\right) \triangleq ||\mathbf{U}_k \mathbf{U}_k^H - \mathbf{A}\mathbf{A}^H||_2$ denotes the projection 2-norm distance between $\mathbf{U}_k$ and $\mathbf{A}$. Geometrically, each subspace spanned by $\mathbf{U}_k$ is a point on a Grassmann manifold. Hence, the optimal aggregation beamformer derived from P1 is the centroid of the set of such points corresponding to $\{\mathbf{U}_k\}$ on a Grassmann manifold. Instead of adopting iterative algorithms to find $\mathbf{A}^*$, we use $d_{\text{PF}}\left(\mathbf{U}_k, \mathbf{A}\right) \triangleq ||\mathbf{U}_k \mathbf{U}_k^H - \mathbf{A}\mathbf{A}^H||_\text{F}$ to approximate $d_{\text{P2}}\left(\mathbf{U}_k, \mathbf{A}\right)$ in the objective function to approach closed-form solution, given $d_{\text{P2}}\left(\mathbf{U}_k, \mathbf{A}\right) \approx d_{\text{PF}}\left(\mathbf{U}_k, \mathbf{A}\right)$ if the *principle angle* (between two subspaces) is small [21]. Let $\mathbf{G} \in \mathbb{C}^{M_r \times M_r}$ defined as

$$\mathbf{G} = \sum_{k=1}^{K} \mathbf{U}_k \mathbf{U}_k^H,$$

then the optimal solution (for the approximated objective function) $\mathbf{A}^*$ is given by the first $N$ principal eigen-vectors of $\mathbf{G}$. In detail, let $\mathbf{G} = \mathbf{V}_G \Sigma_G \mathbf{V}_G^H$ be the SVD of $\mathbf{G}$, the *aggregation beamformer* is designed as

$$\mathbf{A}^* = \left[\mathbf{V}_G\right]_{:,1:N}.$$

## 4. Data Preprocessing for Edge Learning

### 4.1 Recent Trends and Opportunities

Data is usually preprocessed prior to learning for the purpose of compression or making learning faster and more effective. For centralized edge learning, typical data preprocessing operations include feature extraction, principle component analysis, linear discriminant analysis, motion representation and low-dimensional representation [22]. For federated edge learning, data are used to update a local model or compute a stochastic gradient, both of which are then transmitted for updating the global model. For wireless transmission, it is also necessary to pre-process data prior to transmission. Typical preprocessing includes source encoding, channel encoding, multi-antenna precoding, interference avoidance, and channel pre-equalizaiton [23]. Their purposes are either efficient compression (source encoding) or protection against channel hostility (other operations). Based the new "computation-communication integration" approach, the data preprocessing for machine learning and wireless transmission can be integrated to minimize the communication overhead and negative effects of channel hostility on the edge learning performance. This establishes the **new principle** for designing data preprocessing for edge learning. There are some recent results based on this principle. For example, data for training a classifier model is encoded into subspace matrices so that they can be transmitted using the channel blind non-coherent MIMO modulation. Such encoding is shown to significantly improve robustness for edge learning w.r.t. conventional transmission schemes in the presence of fast fading.

Despite some advancements, the area of data preprocessing for edge learning is still large open with many promising opportunities. Some are described as follows.

- **Gradient quantization:** Existing *vector quantization* (VQ) algorithms target low-dimensional data and adopt a generic error measure such as *mean squared erro*r (MSE) [24]. For federated edge learning, the high dimensionality of stochastic gradients and their error being measured in terms of direction render the existing VQ algorithms inefficient or even ineffective [25]. This calls for the development of VQ algorithms specialized for gradients to exploit the full high-dimensional VQ gain. A design example from recent work is discussed in the sequel.

- **Communication efficient feature selection**: Selecting useful features from training data not only reduces communication overhead but also accelerates learning. While there exist many methods for feature selection from centralized data [26], the operation is challenging for distributed data. From the communication perspective, designing distributed feature selection should also attempt to minimize communication overhead besides ensuring learning performance.

- **Channel adaptive feature-extraction:** Learning performance tends to improve if more features of the dataset are used. However, transmitting more features requires more radio resources or else incurs longer latency due to more communication load. For edge learning, given time-varying channel capacity, it is necessary to balance the communication load and learning performance by adapting feature extraction to the channel state. In other words, the number of features should be adjusted according to the channel capacity or state in a similar way as adaptive modulation.

- **Channel adaptive gradient compression**: For federated learning, stochastic gradients exhibit sparsity allowing compression by magnitude-based truncation of coefficients [8]. More aggressive truncation reduces communication load but increases the degradation of learning performance. This motivates the adaptation of the gradient truncation level to the channel state to balance communication efficiency and learning performance.

- **Motion data encoding**: A known result is that a typical motion can be represented by a sequence of subspaces, which is translated into a trajectory on a Grassmann manifold [27]. How to encode a motion dataset for both efficient communication and machine learning is an interesting topic for edge learning. Based on the proposed design principle, one possible solution is to encode the whole dataset into a smaller one while preserving the trajectory information. Then, by uploading the smaller dataset, the original trajectory could be recovered by interpolation. The key design challenge relies on how to choose the size and components of the smaller dataset so as to make it representative enough for the original one given the learning latency constraint.

## 4.2 Design Example

In this sub-section, considering federated edge learning, we illustrate the design of high-dimensional stochastic gradient quantization via an example. The example, *hierarchical gradient quantization,* is based on recent work in [28]. The design provides a practical method for quantizing high-dimensional gradients. It successfully reduces the communication overhead as compared to the state-of-the-art scheme, called signSGD [25], while achieving similar learning performances.

Consider the federated edge learning system in Fig. 6 where each edge device computes and quantizes a stochastic gradient and then transmits the quantized gradient to the edge server for aggregation for the purpose of updating the global AI model. Consider an arbitrary communication round. The edge server broadcasts the global model under training to all edge devices. Based on the received model each device computes the stochastic gradient , denoted as $g \in \mathbb{R}^{\text{Dim} \times 1}$, by differentiating the local loss function. To reduce the communication overhead, each device quantizes the gradient prior to transmission. To facilitate the exposition, define the normalized stochastic gradient $f = g/\|g\|$ and the norm of the stochastic gradient $\rho = \|g\|$. For tractability and as verified by experiments, it is reasonable to assume that the normalized stochastic gradient is uniformly distributed on the Grassmann manifold, referring to the space of normalized vectors and hence a hypersphere. A natural approach of quantizing a high-dimensional gradient vector is to divide it into low-dimensional blocks, each of which is quantized using a generic VQ algorithm. This approach faces two issues: 1) fails to exploit the high-dimensional VQ gain by jointly quantizing all blocks; 2) the generic distortion measure (e.g., MSE) is unsuitable for a gradient that aims at conveying information on descending direction.

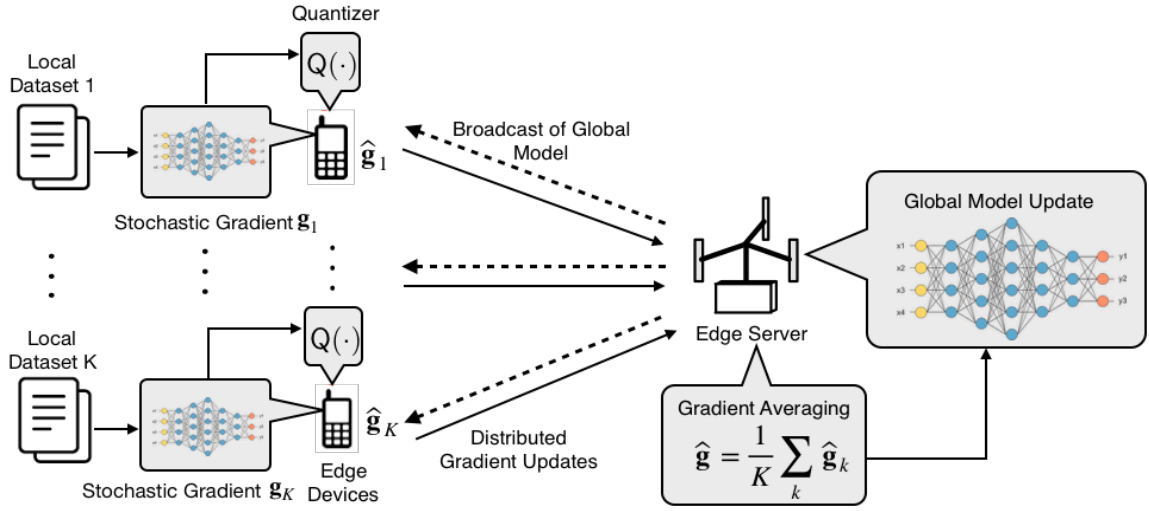

*Fig. 6. Federated edge learning system with quantized gradient transmission.*

The issues can be addressed by the following hierarchical quantization framework proposed in [28]. The framework quantizes the gradient norm with a scalar codebook and the normalized stochastic gradient with two low-dimensional Grassmannian codebooks. This is motivated by the suitability of such a codebook for quantizing a vector that contains directional information [29]. To reduce the quantization complexity, the normalized gradient $f$ is decomposed into $M$ blocks of length $L$, i.e. $f^T = [v_1^T, \cdots, v_M^T]$, where $v_i$ is referred to as the *i*-th *block gradient*. Define the *normalized block gradient* $s_i = v_i/\|v_i\|$ and the hinge vector $h = [h_1, \cdots, h_M]^T$ with $h_i = \|v_i\|$. It follows that both the normalized block gradients and the hinge vector have unit norm and each can be quantized using a Grassmannian codebook. The block gradients admit low-dimensional VQ while the hinge vector assemble the block gradients to yield the quantized gradient under the criterion of minimum distortion in descending direction [28]. In addition, one can show that if the normalized gradient is isotropic, the normalized block gradients are also isotropic. Given the above decomposition and properties, a hierarchical quantization framework is illustrated in Fig. 7. The procedure of **Hierarchical Gradient Quantization** is summarized as follows [28]:

- Decompose a stochastic gradient into a gradient norm, a set of normalized block gradients, and a hinge vector;

- Quantize the gradient norm using a $B_\rho$-bit scalar and uniform quantizer;
- Quantize each normalized block gradient using a $B_s$-bit isotropic Grassmannian codebook designed by a *line packing* algorithm (see e.g., [29]);
- Quantize the hinge vector using a $B_h$-bit Grassmannian codebook with positive coefficient elements and designed using the *Lloyd algorithm (see e.g., [29])*.

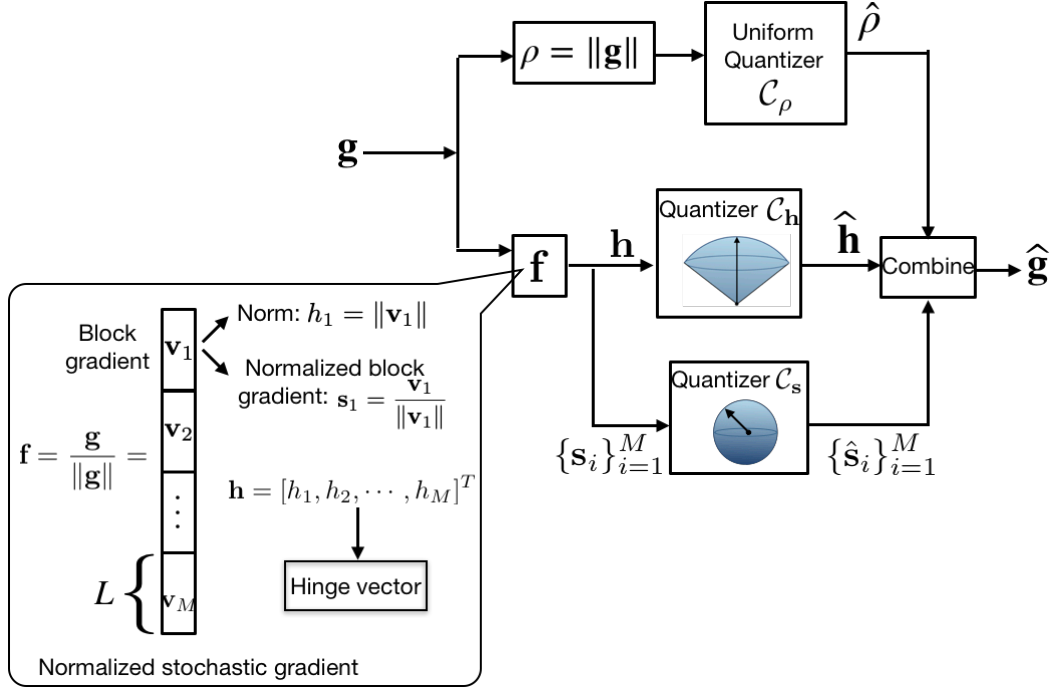

*Fig. 7. Hierarchical gradient quantization.*

After hierarchical quantization, all bits are transmitted from the device to the server. With the prior knowledge on the quantizer codebooks, the server can compute the quantized gradient $\hat{g}$ by table-lookup.

The performance of the hierarchical gradient quantization framework is compared with that of the state-of-the-art signSGD by real experiments. The benchmarking scheme quantizes gradient coefficients using a binary scalar quantizer and is shown to achieve the same order of convergence speed as in the ideal case. The experiment is set up as follows. Consider a federated learning system with one edge server and 100 edge devices. The learning task is to train a classifier model for handwritten-digit recognition using the well-known MNIST dataset that consists of 10 categories ranging from digit "0" to "9" and a total of 60000 labeled training data samples. The classifier model is implemented using a 6-layer *convolutional neural network* (CNN). The total quantization bits for each gradient is allocated using a bit-allocation algorithm in [28]. The communication overhead is measured by the number of quantization bits per gradient coefficient. The effectiveness of the proposed hierarchical quantization scheme is evaluated by benchmarking against signSGD and SGD with unquantized gradients. The curves of the test accuracy versus the number communication rounds are plotted in Fig. 8. Several observations can be made as follows. First, using fewer bits, i.e., 0.5 bit/coefficient, the performance of the proposed scheme is comparable to signSGD, which uses 1 bit/coefficient. This demonstrates the effectiveness of hierarchical quantization in harnessing the high-dimensional VQ gain.

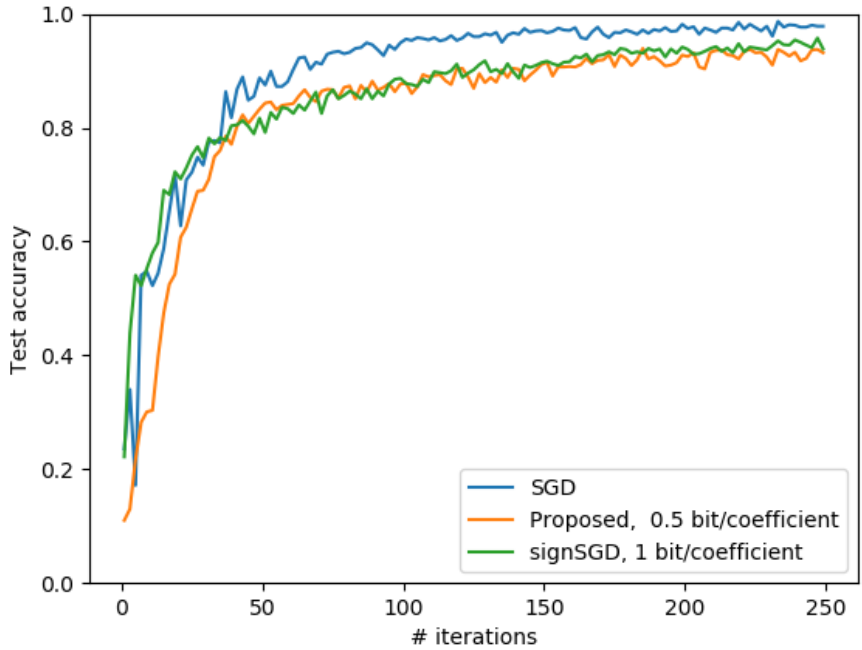

*Fig. 8. Performance comparison of signSGD and the proposed scheme.*

# 5. Radio Resource Management for Edge Learning

## 5.1 Recent Trends and Opportunities

Classic techniques for *radio resource management* (RRM) are designed for radio access networks to provide access to many subscribers. Their objective is to maximize the sum rate under constraints on subscribers' *quality-of-service* (QoS). With this objective, many schemes have been proposed for allocating the finite radio resources (time, spectrum, space or code domain) to subscribers including TDMA [30], OFDMA [31, 32], SDMA [33], and *code-division multiple-access* (CDMA) [34]. However, these schemes are inefficient for edge learning and cannot overcome its communication bottleneck. The main reason is that the objective of rate maximization is not aligned with that of edge learning that aims at efficient intelligence acquisition. This calls for the development of a new class of RRM techniques for communication efficient edge learning based on the following **principle**: radio resources should be allocated to speed up the convergence of learning and improve the accuracy of learning.

The research trends differ for centralized and federated edge learning. For centralized edge learning, the main trend on RRM is to evaluate the importance of distributed data samples for learning and allocate radio resources accordingly to acquire important samples. One data-importance metric for data classification is **data uncertainty** [12]. It measures how confident a data sample can be correctly classified using a classifier model under training. There exist different measures of data uncertainty such as entropy or the distance to a classification boundary. When considering a batch of samples, the total informativeness depends not only on uncertainty of individual samples but also on their **diversity**, referring to non-overlapping information. This yields another data-importance metric, data diversity [13]. The basic idea of **data-importance aware** RRM for centralized edge learning is to design schemes that allocate radio resources to devices for data uploading by jointly considering their data importance measures and channel states. The concept can be extended to federated edge learning though it does not involve direct data uploading. The main idea is to treat the local models/gradients as data, and evaluate their importance levels for model training, and account for such importance in RRM. A suitable importance measure for a local model is *model variance* with respect to the global model; that for a local stochastic gradient is *gradient divergence* that reflects the level of changes on the current gradient update with respect to the previous one. Thereby, radio resources are allocated to devices by jointly considering their model variances (or gradient divergences) together with channel states. Another trend of RRM for federated edge learning targets a key operation in federated edge learning, *synchronous updating*, where the acquisition of local updates are synchronized so that they can be averaged for updating the global model. The main issue addressed by RRM is to avoid a device with slow computing or a weak wireless link becoming a performance bottleneck.

The area of RRM for edge learning is in a beginning stage but attracts strong interests. One can find many promising opportunities in the area. Some are described as follows.

1. **Data-importance aware RMM for centralized edge learning**: A multiuser edge learning system features two kinds of multiuser diversity [35]. One is multi-user channel diversity, referring to independent fading in multiuser channels. The other is multi-user data diversity, referring to heterogeneous importance levels across multiuser datasets [11]. Exploiting multiuser channel diversity for throughput gain is a main theme in conventional scheduling design [36]. In the context of edge learning, a new direction for RRM is to jointly exploit both types of diversity so as to simultaneously maximize the communication efficiency and learning performance. Such designs are required to balance two conflicting goals. One is to maximize the channel capacity or reliability by scheduling devices with the best channels. The other is to select devices with most informative data so as to maximize model convergence rate. They are both important for communication efficient edge learning and thus need to be balanced. As a result, designs should select devices with both relatively reliable channels and data that are sufficiently informative. A design example is provided in the sequel. Other than scheduling, importance aware RRM can be also applied to other dimensions such as power control and spectrum allocation.

2. **Data-importance aware RMM for federated edge learning**: The basic principle is similar to that for the preceding topic. The RRM for federated edge learning also aims at simultaneously exploiting two types of multiuser diversity: one is *multi-user channel diversity* as before and the other is *multiuser update diversity* (in terms of local model variance or gradient divergence). In addition, RRM design for the current case can also leverage a unique *communication-computation tradeoff* that increasing local computation load (the number of local model training rounds) can be traded for reduced communication overhead (the number of communication rounds for updating the global model) [6]. This makes it necessary for RRM designs to account for the heterogeneity in devices' computation capacities as well as their computation power consumption.

3. **Data diversity aware RRM**: Multi-antenna (or broadband) communication are widely adopted in modern communication systems such as LTE to spatial (or frequency) multiplexed parallel data streams [37]. This enables data samples to be acquired by an edge server in batches. Consider selecting a batch of distributed data samples from multiple devices. Besides multiuser channel states and data uncertainty, it is important to consider the diversity of the samples within the same batch. One single measure that integrates both aspects is called *Fisher information matrix* [38]. Such a measure can be combined with the channel state in an intelligent way to design a new scheduling metric to achieve both communication efficiency and data diversity plus uncertainty for accelerating learning.

4. **Energy efficient RRM**: For federated learning with synchronized updates, the latency per round depends on the communication and computation latency of individual devices. On one hand, communication latency of a device can be controlled by allocating more radio resources. On the other hand, the computation latency can be reduced by accelerating the computing speed, which, however, increases the energy consumption [39]. Considering the two types of latency control gives rise to a problem of RRM for simultaneous energy consumption and latency reduction. Formulating and solving such multi-objective optimization problems give rise to a class of new energy efficient RRM techniques (see e.g., [40]).

## 5.2 Design Example

Based on the work in [35], this example targets a centralized edge learning system as shown in Fig. 3(b). There are one server and $K$ devices. The learning task is to train a classifier model at the server. The devices access the channel using TDMA to upload local training data samples to the server. The system operations in each communication round are described as follows. First, the edge server broadcasts the current global model to all devices. Second, each device uses the received model to evaluate the importance of a randomly selected data sample and reports the *data importance indicator* (DII), denoted as $I_k$ for the $k$-th device, to the server. Third a device is scheduled for transmission based on the criterion of maximum DII. Analog modulation that is known to be efficient for multimedia transmission [41] is used to modulate transmitted data. Last, the server uses the received data sample to update the model and then broadcasts the updated model, completing the communication round.

The learning model is elaborated as follows. The classifier is based on a binary soft-margin SVM model. Before wireless data acquisition, a coarse initial classifier is available at the server so that data-importance evaluation can be performed at the beginning using the model. The model is refined progressively in the training process. The importance of a noisy received data sample can be measured using its *expected uncertainty*. Since the uncertainty is higher for a sample nearer to the decision boundary and vice versa, the expected uncertainty can be translated to the expected distance between the data sample and the current decision boundary of the classifier model.

Targeting the scenario where labelling is costly, the data samples in the devices are assumed to be unlabelled. And the label is generated for the selected data after transmission to the server by recruiting a labeller. We should first quantify the effect of channel fading and noise on the expected received data samples at the server. Let $\text{SNR}_k$ and $\mathbf{x}_{k,n}$ denote the SNR and the $n$-th data sample of the $k$-th device, and $d(\mathbf{x}_{k,n})$ denote the distance between $\mathbf{x}_{k,n}$ and the decision boundary, respectively. The DII of the selected data sample in the $k$-th device, refers to the sample with the maximum expected uncertainty among the local data. In [35], the DII is derived as

$$I_k = -\frac{1}{\text{SNR}_k} + \max_{n \in \mathcal{N}_k} \mathcal{U}_\text{d}\left(\mathbf{x}_{k,n}\right), \tag{1}$$

where $\mathcal{N}_k = \{1,2,\cdots,N\}$ represents the sample index set, and $\mathcal{U}_\text{d}(\cdot)$ is a distance-based uncertainty measure. One can observe that the derived DII includes both data uncertainty (the last term) and the channel quality (the SNR term) in a simple addition form. The DII being a monotone increasing function of SNR is due to the fact that the channel fading and noise tend to degrade the importance of data samples by making their distances to the decision boundary more likely to be larger than smaller. In other words, channel distortion degrades the learning performance as expected. Based on the DII in (1), the scheduling scheme for binary SVM is to select the device with largest DII. To be specific, the edge server schedules device $k^*$ for data transmission if

$$k^* = \arg\max_k \left\{ -\frac{1}{\text{SNR}_k} + \max_{n \in \mathcal{N}_k} \mathcal{U}_\text{d}\left(\mathbf{x}_{k,n}\right) \right\}. \tag{2}$$

The design can be easily extended to a general classifier (such as CNN) by replacing the distance based uncertainty measure with a general measure [35]. The scheduling metric, namely the DIII in (1), shows that both multiuser channel diversity (first term) and multi-user data diversity (second term) should be exploited for learning performance improvement.

The performance of data-importance scheduling is evaluated against that of conventional designs as follows. The first baseline scheme, channel-aware scheduling, only exploits the multiuser channel diversity; the other, data-aware scheduling, only leverages the multiuser data diversity. The experiment has the following settings. There are $K = 10$ edge devices in the system. The transmission budget for the binary SVM learning task is $T = 100$ channel uses. Each channel use is for transmitting a single data sample. Rayleigh fading channels with unit variance are considered with the average transmit SNR=15 dB. The well-known MNIST dataset also used in previous experiments is adopted for training. The test accuracies of models trained using different schemes are compared in Fig. 9. One can observe that importance-aware scheduling can achieve signifiant improvement in test accuracy of about 5% over channel-aware scheduling and of about 8% over data-aware scheduling. Moreover, the model convergence of the new design is faster too.

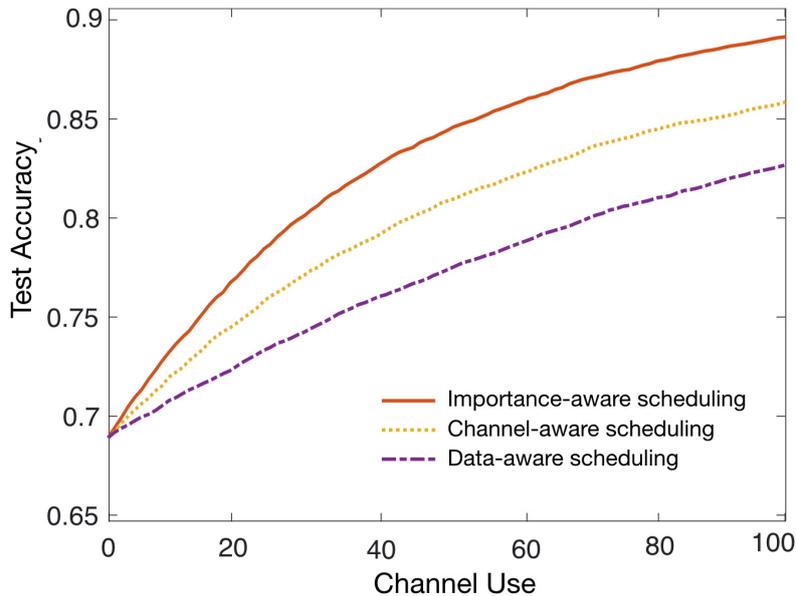

*Fig. 9. Learning performance of importance-aware scheduling versus channel-aware and data-aware scheduling.*

# 6. Concluding Remarks

The past decades have seen breath-taking advancements in both wireless communication and computer science, which have occurred in parallel with few cross paths. However, the 5G-and-beyond vision of ubiquitous edge intelligence calls for new technologies seamlessly integrating two disciplines to realize large-scale AI for solving grand problems our society is facing ranging from autonomous transportation to digital agriculture. In view of the trend, there is no doubt that edge learning will shift the paradigms of both communication and computing. Relevant technologies will play an important role in the future 6G era. Hopefully, this article has succeeded in providing readers a comprehensive introduction of communication efficient edge learning and new comers useful guidelines for researching in this new area.